\documentclass[epj]{svjour}
\usepackage{amsmath,amssymb}
\usepackage[square,comma,numbers,sort&compress]{natbib}
\usepackage{graphicx}
\usepackage{pstricks}

\newcommand{\ket}[1]{|#1\rangle}
\newcommand{\bra}[1]{\langle #1|}
\newcommand{\braket}[2]{\langle #1|#2\rangle}
\newcommand{\ketbra}[2]{|#1\rangle\!\langle #2|}

\newcommand{\idty}{\mathcal{I}}
\newcommand{\abs}[1]{\ensuremath{\vert #1 \vert}}

\newcommand{\realcomment}[1]{}

\DeclareMathOperator{\CNOT}{CNOT}
\DeclareMathOperator{\tr}{tr}
\begin{document}
\title{Stabilization of quantum information by combined dynamical decoupling and detected-jump error correction}

\author{Daniel Geberth\inst{1} \and Oliver Kern\inst{1}  \and Gernot Alber\inst{1}
\and Igor Jex\inst{2}
}                     
\offprints{}          
\institute{Institut f\"ur Angewandte Physik,
Technische Universit\"at Darmstadt,
64289~Darmstadt, Germany \and
Department of Physics,
FJFI \v CVUT,
B\v rehov\'a 7,
115 19 Praha 1 - Star\'e  M\v{e}sto, Czech Republic
}
\date{Received: date / Revised version: date}
%
\abstract{
Two possible applications of random decoupling are discussed.
Whereas so far decoupling methods have been considered merely for quantum memories,
here it is demonstrated that random decoupling is also a convenient tool for stabilizing quantum algorithms.
Furthermore, a decoupling scheme is presented which involves a random decoupling method 
compatible with detected-jump error correcting quantum codes. With this combined error correcting strategy
it is possible to stabilize
quantum information against both spontaneous decay and static imperfections of a
qubit-based quantum information processor in an efficient way.
} 
\titlerunning{Combined dynamical decoupling and detected-jump error correction}
\maketitle

\section{Introduction}
Quantum information processing of many-qubit systems requires 
efficient methods for
protecting arbitrary logical quantum states 
against unwanted environmental interactions or uncontrolled inter-qubit couplings.
So far, two strategies have been developed for achieving this goal, namely dynamical decoupling and redundancy-based
quantum error correction. Dynamical decoupling \cite{VKL99} is based on the main idea of suppressing unwanted influences on the
quantum dynamics by appropriately applied deterministic or stochastic external forces. A disadvantage of this approach is that
typically 
residual errors are left. An advantage of this method of error suppression is that 
the complete physically available Hilbert space can be used for
logical information processing. In redundancy-based error correction \citep[see][chapter 10]{nielsenchuang} only a subspace of the physically available
Hilbert space can be used for logical information processing. This subspace has to be 
selected in such a way that 
possible errors can be corrected perfectly by appropriate control measurements and recovery operations. 
Thus, this latter type of error correction typically requires a sufficiently large number of redundant qubits which cannot be used
for logical purposes. This represents a serious disadvantage if 
only quantum information processors  with small numbers of qubits are available.
Thus, in view of nowadays technology it is desirable to develop error correcting strategies which
combine
the positive advantages of both approaches in order to achieve the highest possible degree of error suppression
with the smallest possible number of redundant qubits.
Combined methods in which environmental interactions are corrected by redundancy-based quantum error correction and in which
uncontrolled inter-qubit
couplings within a quantum information processor
are suppressed by dynamical decoupling may offer interesting perspectives for the stabilization of few-qubit
quantum information processors against
unwanted errors.

In the following such a combined error correction scheme is presented.
The main problem which has to be overcome in this context 
is compatibility. 
One has to ensure that the chosen dynamical decoupling method does not leave the code space of the
redundancy-based quantum error correction method.
If this compatibility cannot be guaranteed uncorrectable errors may arise.
Furthermore, in practical applications one is also interested in minimizing the residual errors originating from
errors during recovery operations, for example, and from the dynamical decoupling method itself.
Currently, no universal solution is available for this compatibility and optimization problem. 

Here, first results and solutions of these
compatibility and optimization problems
are presented for a particular combined error correcting method which is capable of
correcting a special case of environmentally-induced decoherence, namely
spontaneous decay of qubits. Such decay processes may arise from spontaneous emission of photons,
for example. In our subsequent discussion we concentrate on the special but important case that
spontaneous decay processes affecting different qubits are statistically independent. This implies  that in principle
not only error times but also error
positions can be determined by observation of the individual qubits. In the context of
photon-induced spontaneous decay this case is realized whenever the wave lengths of spontaneously emitted
photons are significantly smaller than the spatial separations
between adjacent qubits. Under these circumstances detected-jump error correcting quantum codes \cite{ABC01,alber2}
can be used for an efficient redundancy-based correction of resulting errors.
As an additional source of errors we consider uncontrolled inter-qubit couplings within a quantum information
processor which can be modeled by Heisenberg-type couplings with unknown parameters.
It has been demonstrated recently that
such errors can be suppressed significantly with the help of embedded dynamical decoupling methods in which
a deterministic dynamical decoupling scheme is embedded into a stochastic one \cite{combi}.
The deterministic decoupling scheme leads to a significant error suppression
for short times whereas the stochastic decoupling method guarantees that the residual error-induced exponential decay of the fidelity of 
any quantum state is linear-in-time only and not quadratic as in the case of a pure deterministic scheme \cite{parec,VK05}.

This paper is organized as follows.
In Sec. 2 basic concepts of deterministic, stochastic, and embedded dynamical decoupling methods are summarized briefly.
In Sec. 3 the error suppressing potential of stochastic dynamical decoupling methods
is studied quantitatively with the help of the entanglement fidelity. This latter quantity
is identical to the mean fidelity for large numbers of qubits.
Whereas previously derived rigorous lower bounds \cite{VK05} tend to overestimate the fidelity decay of a quantum memory significantly
this mean fidelity yields reliable quantitative estimates on the mean accuracy of achievable stabilization properties
for both quantum memories and quantum computations.
In Sec. 4, finally, a new combined error correcting scheme is introduced in which a compatible embedded dynamical decoupling scheme
is applied within a detected-jump correcting quantum code space.
This way static random imperfections and spontaneous decay processes can be suppressed simultaneously.

\section{Dynamical Decoupling - Basic Concepts}\label{sec::DD}
Let us consider a quantum system $S$ whose Hilbert space $\mathcal{H}_S$ is of dimension $N$.
Typically $S$ represents a quantum register containing $n_q$ qubits so that $N = \dim \mathcal{H}_S = 2^{n_q}$.
The time evolution $U(t)$ of $S$ is due to a time dependent Hamiltonian $H(t) = H_0 + H_c(t)$ which is the sum of a static term we would like to suppress
(this term may represent imperfections in a quantum computer, for example),
and a time dependent term which represents the control operations we are able to apply.
Dynamical decoupling tries to achieve a suppression of $H_0$.

According to the Schr\"odinger equation our system evolves in time as
\begin{equation}\label{eq::u}
U(t) = \mathcal{T} \exp \left( -i\int_0^t H(\tau) d\tau/\hbar \right).
\end{equation}
Analogous to \eqref{eq::u}
let us denote 
the time evolution due to $H_c(t)$ alone by $U_c(t)$, i.\,e.
\begin{equation}
U_c(t) = \mathcal{T} \exp \left( -i\int_0^t H_c(\tau) d\tau/\hbar \right).
\end{equation}
($\mathcal{T}$ denotes the time ordering operator.)
The time evolution in the so called toggled frame $\tilde{U}(t) = U_c^\dagger(t) U(t)$ is determined by the
Schr\"odinger equation
\begin{equation}
i\hbar  \frac{d\tilde{U}(t)}{dt} = \tilde{H}(t) \tilde{U}(t), \qquad \tilde{H}(t)=U_c^\dagger(t) H_0 U_c(t).
\end{equation}
In a simplified scenario the time dependence of $H_c(t)$ involves sequences of $\delta$-pulses
applied at times $t_i = i\Delta t,~i\in \mathbb{N}$.
This time evolution gives rise to
so-called unitary bang-bang pulses of the form $g_{f(i)} g_{f(i-1)}^\dagger$, whereas the $i$-th pulse is the product of elements
$g$ which are chosen from a set of unitary operators $\mathcal{G}$ according to a certain selection rule $f(i)$.
For the corresponding
time evolution due to $H_c(t)$ we obtain
\begin{equation}
U_c(t)=g_{f(i)},  \quad (i-1)\Delta t \leq t \leq i\Delta t
\label{UC}
\end{equation}
and the Hamiltonian $\tilde{H}(t)$ of the toggled frame becomes piecewise constant, i.e.
\begin{equation}
\tilde{H}(t)=g_{f(i)}^\dagger H_0 g_{f(i)}, \quad (i-1)\Delta t \leq t \leq i\Delta t.
\end{equation}
In the following several selection rules $f(i)$ and their advantages and disadvantages will be discussed.
They give rise to periodic (or cyclic) decoupling, random decoupling, and embedded decoupling schemes.

\begin{figure*}
\small%
\begin{center}
\begin{pspicture}(0,0)(15,-5.6)
%
\psline[linewidth=1pt,arrowsize=7pt]{<-}(0.75,-1.)(15,-1.)
\rput[t](0.25,-0.5){\small (a)}
\rput[t](0.23,-1.15){\footnotesize time}
\psline(15,-1.1)(15,-0.9)\rput[t](15,-1.3){\small $0$}\rput[b](15,-0.8){\small $\hat{g}_1$}
\psline(13,-1.1)(13,-0.9)\rput[t](13,-1.3){\small $\Delta t$}\rput[b](13,-0.8){\small $\hat{g}_2 \hat{g}_1^\dagger$}
\psline(11,-1.1)(11,-0.9)\rput[t](11,-1.3){\small $2\Delta t$}\rput[b](11,-0.8){\small $\hat{g}_3 \hat{g}_2^\dagger$}
\psline[linewidth=1pt,linestyle=dashed,linecolor=white](10,-1.)(9,-1.)
\psline(8,-1.1)(8,-0.9)\rput[t](8,-1.3){\small $T_c = n_c\Delta t$}\rput[b](8,-0.8){\small $\hat{g}_1 \hat{g}_{n_c}^\dagger$}
\psline(6,-1.1)(6,-0.9)\rput[t](6,-1.3){\small $T_c+\Delta t$}\rput[b](6,-0.8){\small $\hat{g}_2 \hat{g}_1^\dagger$}
\psline(4,-1.1)(4,-0.9)\rput[t](4,-1.3){\small $T_c+2\Delta t$}\rput[b](4,-0.8){\small $\hat{g}_3 \hat{g}_2^\dagger$}
\psline[linewidth=1pt,linestyle=dashed,linecolor=white](3,-1.)(2,-1.)
\psline[linewidth=1pt,arrowsize=7pt]{<-}(0.75,-3.)(15,-3.)
\rput[t](0.25,-2.5){\small (b)}
\rput[t](0.23,-3.15){\footnotesize time}
\psline(15,-3.1)(15,-2.9)\rput[t](15,-3.3){\small $0$}\rput[b](15,-2.8){\small $\hat{r}_1$}
\psline(13,-3.1)(13,-2.9)\rput[t](13,-3.3){\small $\Delta t$}\rput[b](13,-2.8){\small $\hat{r}_2 \hat{r}_1^\dagger$}
\psline(11,-3.1)(11,-2.9)\rput[t](11,-3.3){\small $2\Delta t$}\rput[b](11,-2.8){\small $\hat{r}_3 \hat{r}_2^\dagger$}
\psline(9,-3.1)(9,-2.9)\rput[t](9,-3.3){\small $3\Delta t$}\rput[b](9,-2.8){\small $\hat{r}_4 \hat{r}_3^\dagger$}
\psline(7,-3.1)(7,-2.9)\rput[t](7,-3.3){\small $4\Delta t$}\rput[b](7,-2.8){\small $\hat{r}_5 \hat{r}_4^\dagger$}
\psline[linewidth=1pt,linestyle=dashed,linecolor=white](3,-3.)(2,-3.)
\psline[linewidth=1pt,arrowsize=7pt]{<-}(0.75,-5.)(15,-5.)
\rput[t](0.25,-4.5){\small (c)}
\rput[t](0.23,-5.15){\footnotesize time}
\psline(15,-5.1)(15,-4.9)\rput[t](15,-5.3){\small $0$}\rput[b](15,-4.8){\small $\hat{g}_1 \hat{r}_1$}
\psline(13,-5.1)(13,-4.9)\rput[t](13,-5.3){\small $\Delta t$}\rput[b](13,-4.8){\small $\hat{g}_2 \hat{g}_1^\dagger$}
\psline(11,-5.1)(11,-4.9)\rput[t](11,-5.3){\small $2\Delta t$}\rput[b](11,-4.8){\small $\hat{g}_3 \hat{g}_2^\dagger$}
\psline[linewidth=1pt,linestyle=dashed,linecolor=white](10,-5.)(9,-5.)
\psline(8,-5.1)(8,-4.9)\rput[t](8,-5.3){\small $T_c = n_c\Delta t$}\rput[b](8,-4.8){\small $\hat{g}_1 \hat{r}_2 \hat{r}_1^\dagger \hat{g}_{n_c}^\dagger$}
\psline(6,-5.1)(6,-4.9)\rput[t](6,-5.3){\small $T_c+\Delta t$}\rput[b](6,-4.8){\small $\hat{g}_2 \hat{g}_1^\dagger$}
\psline(4,-5.1)(4,-4.9)\rput[t](4,-5.3){\small $T_c+2\Delta t$}\rput[b](4,-4.8){\small $\hat{g}_3 \hat{g}_2^\dagger$}
\psline[linewidth=1pt,linestyle=dashed,linecolor=white](3,-5.)(2,-5.)
\end{pspicture}
\end{center}
\normalsize
\caption{Schematic representation of:
(a) periodic dynamical decoupling (\mbox{\textsf{PDD}}) $g \in \mathcal{G}$,
(b) naive random decoupling (\mbox{\textsf{NRD}}) $r \in \mathcal{G}$,
(c) embedded decoupling (\mbox{\textsf{EMD}}) $g \in \mathcal{G}$ and $r \in \mathcal{G}'$.
The time evolution between subsequent instantaneously applied unitary operations is assumed to be governed by a perturbing Hamiltonian $\hat{H}_0$.
\label{fig::schemes}}
\end{figure*}
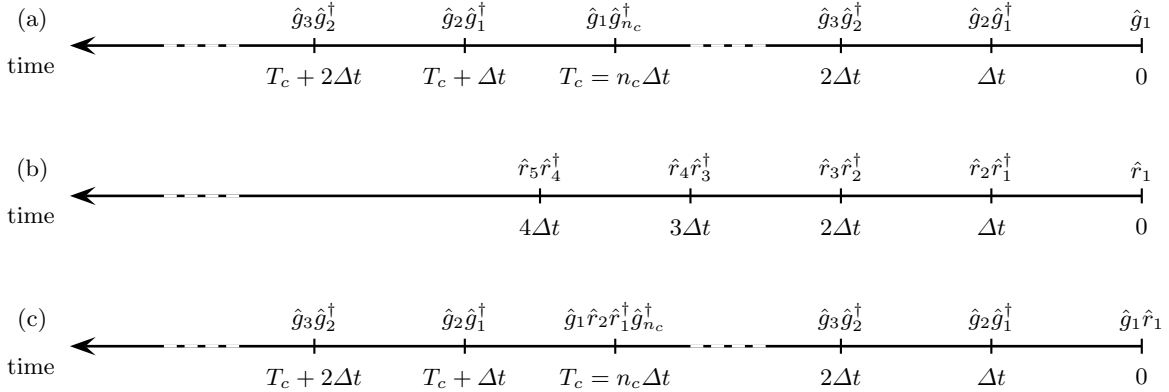

\subsection{Periodic Decoupling}
Let us chose $f(i) = ( i-1 \bmod{n_c} ) + 1$ so that $U_c(t)$ reduces to a set of $n_c$ periodically repeating unitary transformations
$g_{f(i)}\in \mathcal{G}$ (compare with \ref{fig::schemes}a).
This leads to a cyclic unitary control transformation, i.e.
\begin{equation}
U_c(t+T_c) = U_c(t), \quad T_c = n_c \Delta t.
\end{equation}
We can use average Hamiltonian theory (AHT) \cite{nmrbook} to express the unitary time evolution operator
of the toggled frame at integer numbers $n$ of cycles as
\begin{equation}
\tilde{U}(n T_c) = \exp \left( -i \overline{H} n T_c/\hbar \right).
\end{equation}
Thereby,
$\overline{H}$ is given by the Magnus expansion $\overline{H} = \overline{H}{}^{(0)} + \overline{H}{}^{(1)} + \overline{H}{}^{(2)} + \dots$ \cite{nmrbook}.
In the limit of fast control sequences, i.\,e. $T_c \rightarrow 0$ and $M \rightarrow \infty $ for a given time $T = M T_c >0$, it is sufficiently accurate to consider lowest order AHT yielding
\begin{equation}
\overline{H}{}^{(0)} = \frac{1}{T_c} \int_0^{T_c} dt_1 \tilde{H}(t_1)
 = \frac{1}{n_c  \Delta t} \sum_{i=1}^{n_c} g_i^\dagger H_0 g_i \Delta t.
\end{equation}
In order to achieve the main goal of dynamical decoupling, namely a vanishingly small total time evolution in the toggled frame, $\overline{H}{}^{(0)}$ has to vanish.
%
%
This latter requirement can be fulfilled by selecting a set $\mathcal{G}$ of unitary operations, for example,
for which the relation
\begin{equation}
\sum_{g\in \mathcal{G}} g^\dagger H_0 g = 0
\label{Magnus1}
\end{equation}
holds.
Such a set $\mathcal{G}$ of size $n_c=\vert \mathcal{G} \vert$ we call a decoupling set\footnote{On how to construct such a set see e.\,g. \cite{SM01}} for the Hamiltonian $H_0$.
This first order decoupling works perfectly only in the limit of fast control sequences.

\citeauthor{VK05} \cite{VK05} calculated a lower bound for the worst case fidelity
\begin{equation}
F_w(T = nT_c)=\min_{\ket{\psi}}  \vert \bra{\psi} \tilde{U}(T) \ket{\psi} \vert^2
\end{equation}
which is valid for arbitrary periodic dynamical decoupling (\mbox{\textsf{PDD}}\footnote{The abbreviations for the different decoupling methods are adopted from \cite{SV06}.}) sequences.
In particular, they showed
\begin{equation}\label{PDD}
F_w^{\mbox{\scriptsize\textsf{PDD}}} > 1 - \mathcal{O}( T^2 T_c^2 \kappa^4  )
\end{equation}
with $\kappa = \Vert H_0 \Vert_2$
denoting the operator-2 norm of the internal Hamiltonian $H_0$.

A common procedure for obtaining higher than lowest order decoupling is to engineer each cycle in such a way that is it time symmetric.
This can be achieved by traversing an original cycle $\mathcal{G}$ after its completion again backwards so that the relation $\tilde{H}(T_c'-t)=\tilde{H}(t)$ holds with $T_c'=2T_c$.
In such a case all terms of odd orders in the Magnus expansion vanish and one obtains the improved bound \cite{SV06}
\begin{equation}\label{SDD}
F_w^{\mbox{\scriptsize\textsf{SDD}}} > 1 - \mathcal{O}( T^2 T_c^4 \kappa^6 ).
\end{equation}
Thereby, \mbox{\textsf{SDD}} stands for symmetric dynamic decoupling.

A disadvantage of periodic (or cyclic) decoupling is the quadratic time dependence of the fidelity decay and its dependence on the length of the sequence.
As a consequence for long times and/or large decoupling sets $\mathcal{G}$ the accumulating errors might increase quickly.

\subsection{Annihilators}
We defined a decoupling set $\mathcal{G}$ for a Hamiltonian $H_0$ as a set of unitary operation $g\in \mathcal{G}$ such that
\begin{equation}
\sum_{g\in \mathcal{G}} g^\dagger H_0 g = 0.
\end{equation}
We will call a decoupling set for which the above equation holds for all traceless Hamiltonians an annihilator.
It was shown in \cite{WRJB02} that such a set contains at least $N^2$ elements.
An annihilator can always be chosen to be also a so called nice error basis, which is a unitary and orthonormal operator basis fulfilling
some additional convenient properties (for a detailed definition see \cite{Knill96a}).
Since the elements of a nice error basis form an irreducible projective representation of an underlying index group Schur's lemma implies the important relation \cite{WRJB02}
\begin{equation}
\frac{1}{\vert \mathcal{G} \vert} \sum_{g\in \mathcal{G}} g^\dagger H_0 g = \frac{ \tr H_0 }{N} \mathcal{I}.
\end{equation}

In our subsequent discussion we will take as annihilator the nice error basis
\begin{equation}\label{eq::pauli}
\mathcal{P} = \{ \mathcal{I},X,Y,Z \}^{ \otimes n_q },
\end{equation}
which consists of tensor products of the Pauli matrices,
\begin{equation}
\mathcal{I} = \left( \begin{array}{cc} 1 & 0 \\ 0 & 1 \end{array} \right),
X = \left( \begin{array}{cc} 0 & 1 \\ 1 & 0 \end{array} \right),
Y = \left( \begin{array}{cc} 0 & -i \\ i & 0 \end{array} \right),
Z = \left( \begin{array}{cc} 1 & 0 \\ 0 & -1 \end{array} \right).
\end{equation}
However, any other annihilator would lead to the same results.

\subsection{Random Decoupling}\label{sec::random}
Let us choose $f(i) = \in_R \{1,2,\dots,n_c\}$,
i.\,e. elements $f(i)$ are selected randomly from the set $\{1,2,\cdots,n_c\}$
giving rise to randomly selected unitary transformations $r_i=g_{f(i)}$ (see \ref{fig::schemes}b).
To calculate the worst case mean fidelity of such a naive random decoupling (\mbox{\textsf{NRD}}) scheme
we consider the average over all realizations of the $r_i$ which we will denote by $\mathbf{E}$, i.e.
\begin{equation}\label{worstcNRD}
F_w(T = n \Delta t)=\min_{\ket{\psi}} \mathbf{E} \vert \bra{\psi} \mathcal{I}^\dagger \tilde{U}(T) \ket{\psi} \vert^2.
\end{equation}
For the moment we call this decoupling scheme 'naive' since there are more sophisticated decoupling methods which will be discussed in the next subsection.
It was shown in \cite{VK05} that this mean fidelity obeys the inequality
\begin{equation}\label{NRD}
F_w^{\mbox{\scriptsize\textsf{NRD}}} > 1 - \mathcal{O}( T \Delta t \kappa^2 ).
\end{equation}

An advantage of \mbox{\textsf{NRD}} methods versus periodic methods is the linear time dependence of the fidelity decay and its independence of the length of the decoupling set $\mathcal{G}$.
However, at the same time the error suppression is weaker (compare the factor $\kappa^2$ of (\ref{NRD}) with the factors $\kappa^4$ of (\ref{PDD}) or $\kappa^6$ of (\ref{SDD})).
As a consequence of this independence it is always possible to draw the random unitary transformations from an annihilator, such as $\mathcal{G}=\mathcal{P}$.
However, choosing a particularly adjusted decoupler $\mathcal{G}$ for a Hamiltonian $H_0$ may offer advantages if one is restricted to particular unitary transformations.

In addition \mbox{\textsf{NRD}} schemes are well suited for stabilizing quantum algorithms against perturbations \cite{parec}.
This aspect will be discussed in detail in section \ref{sec::stabalg}.
In the latter reference we introduced the term Pauli-random-error-correction (PAREC) for the stabilization of a quantum computation using \mbox{\textsf{NRD}} decoupling with $\mathcal{G}=\mathcal{P}$.

\subsection{Embedded Decoupling}
In embedded decoupling (\mbox{\textsf{EMD}}) schemes
two different sets of decoupling operations are used, namely
a set $\mathcal{G}$
which achieves (at least) first order decoupling for a perturbing Hamiltonian $H_0$
and an annihilator $\mathcal{G}'$.

Using decoupling operations in a deterministic way from a set $\mathcal{G}$ for \mbox{\textsf{PDD}} [or \mbox{\textsf{SDD}}]
we obtain an effective time evolution which can be described by $\exp( -i \overline{H} T_c )$. According to (\ref{Magnus1}) the
Magnus expansion yields $\overline{H}{}^{(0)}=0$ for  a \mbox{\textsf{PDD}} scheme [and $\overline{H}{}^{(0)}=\overline{H}{}^{(1)}=0$ for a
\mbox{\textsf{SDD}} scheme].
We can suppress $\overline{H}$ even further by applying a \mbox{\textsf{NRD}} scheme
involving a set of unitary operations $\mathcal{G}'$. Thereby, random unitary operations from this set
are applied inbetween any two \mbox{\textsf{PDD}} or \mbox{\textsf{SDD}}] cycles (compare with figure \ref{fig::schemes}c).

A bound on the worst case fidelity we can obtained by simply taking the \mbox{\textsf{NRD}} bound of \eqref{NRD} and replacing $\Delta t$ by $T_c$ and $\kappa$ by a bound on the norm of $\overline{H}$.
Thus one finds the results \cite{combi}
\begin{equation}
F_w^{\mbox{\scriptsize\textsf{EMD}}} > 1 - \mathcal{O}( T T_c^3 \kappa^4 )
\end{equation}
for an embedded decoupling scheme (\mbox{\textsf{EMD}}) based on an underlying \mbox{\textsf{PDD}} scheme and \cite{SV06}
\begin{equation}
F_w^{\mbox{\scriptsize\textsf{SEMD}}} > 1 - \mathcal{O}( T T_c^5 \kappa^6 )
\end{equation}
for a symmetric embedded decoupling scheme (\mbox{\textsf{SEMD}}) based on an underlying \mbox{\textsf{SDD}} scheme.

Alternatively one may achieve comparably good results by using random path decoupling schemes as discussed in \cite{SV06}.
In these schemes one uses a \mbox{\textsf{PDD}} (or \mbox{\textsf{SDD}}) scheme based on a set $\mathcal{G}$ and in each cycle $j$ the selection rule $f_j(i) = \pi_j( ( i-1 \bmod{n_c} ) + 1 )$ is used. Thereby,
$\pi_j$ is a permutation of the set $\{1,2,\dots,n_c\}$ chosen at random and independently for each cycle.

\section{Stabilizing quantum algorithms against static imperfections by naive random decoupling}\label{sec::stabalg}
As mentioned in section \ref{sec::random} naive random decoupling (\mbox{\textsf{NRD}}) can be used
to stabilize arbitrary quantum algorithms against static imperfections in a rather simple way.
In this section basic properties of this stabilization are explored quantitatively.
As a quantitative measure for the accuracy of stabilization the expectation value of the entanglement fidelity is used which is approximately equal to the average fidelity in the case of high dimensional quantum systems.
Numerical results demonstrating characteristic stabilization properties of the recently proposed PAREC method are presented for the special case of the quantum Fourier transform.
In particular, they are compared with the recently proposed stabilization method of Prosen et al. \cite{Prosen02}.

\subsection{Entanglement fidelity and average fidelity}
The bound \eqref{NRD} was calculated for the worst case fidelity.
Typically this bound overestimate the actual fidelity decay to a high degree.
Therefore, for practical purposes it is more convenient to consider other  fidelity measures, such as the
average fidelity
\begin{equation}\label{Afidelity}
\overline{F}(\mathcal{E}) = \int d\psi \bra{\psi} \mathcal{E}(\ketbra{\psi}{\psi}) \ket{\psi}
\end{equation}
or the entanglement fidelity
\begin{equation}\label{Efideilty}
F_e(\mathcal{E}) = \bra{\phi} (\mathcal{I}\otimes\mathcal{E})(\ketbra{\phi}{\phi}) \ket{\phi}.
\end{equation}
Thereby, the integration involved in the definition of the average fidelity \eqref{Afidelity} has to be  performed over the uniform (Haar) measure on the relevant quantum state space with the normalization $\int d\psi=1$ and $\mathcal{E}$ denotes a trace-preserving quantum operation, i.e. a general completely positive map.
The pure quantum state $\ket{\phi}$ occurring in the definition of the entanglement fidelity is a maximally entangled state between the quantum system under consideration and an ancilla system of the same dimension $N$.
Furthermore, $\mathcal{I}$ denotes the identity operation acting on the ancilla system.
Accordingly, the entanglement fidelity measures the degree to which the entanglement of quantum state is preserved by a quantum operation $\mathcal{E}$.
Apparently, it is independent of the choice of the maximally entangled state
since any two maximally entangled states are related by a unitary acting only on the ancilla.
Both fidelity measures are not independent but are related by \cite{Nielsen02,PBLO04} 
\begin{equation}\label{fidcomp}
\overline{F}(\mathcal{E}) = \frac{N F_e(\mathcal{E}) + 1}{ N+1 } = F_e(\mathcal{E}) + \mathcal{O}\left(\frac{1- F_e(\mathcal{E})}{N}\right).
\end{equation}
Thus, in the case of high dimensional quantum systems, i.e. $N=2^{n_q}\gg 1$, the difference between both measures tends to zero.

In the following we are mainly interested in the entanglement fidelity comparing a unitary operation $U$ and its slightly perturbed version
$U_\delta$. Thus the relevant quantum operation $\mathcal {E}$ involves
a single unitary Kraus operator $K$ which is given by
$K = U^\dagger \cdot U_\delta$.
On the basis of \eqref{fidcomp} in the case of high dimensional quantum systems
the average fidelity is approximately given by the entanglement fidelity 
\begin{equation}
F_e( U^\dagger \circ U_\delta ) = \left \vert \frac{1}{N} \tr \{ U^\dagger U_\delta \} \right \vert^2
\end{equation}
which 
is determined by
the absolute square of a fidelity amplitude
\begin{equation}
 A = \frac{1}{N} \tr \{ U^\dagger U_\delta \}.
\label{fidampl}
\end{equation}
Typically, the evaluation of this fidelity amplitude is much simpler than the direct evaluation of the average fidelity \eqref{Afidelity}.
Therefore, in view of its
close relationship to the average fidelity
our subsequent discussion will
concentrate on the behavior of the entanglement fidelity mainly.

\subsection{Static imperfections and short-time behavior of the fidelity}
Typically,
the fundamental unitary transformation $U$ constituting a quantum algorithm  can be decomposed into a sequence of $n_g$ elementary
one- and two-qubit quantum gates, i.e.
\begin{equation}
U = U_{n_g} \cdot \dots \cdot U_3 \cdot U_2 \cdot U_1.
\end{equation}
Let us assume in our subsequent discussion
that the quantum algorithm under consideration involves $t$ interactions of such a fundamental unitary transformation $U$. Such quantum algorithms
appear in the context of search algorithms, for example \cite{grov1}.
Furthermore, let us focus our attention on the case of static imperfection in which the perturbing influence on such a quantum algorithms
arises from unknown (but fixed and time independent) Hamiltonian coupling $H_0$ between the qubits constituting the quantum information processor. 
In the idealized situation that the controlled elementary quantum gates $U_j,~j=1,...,n_g$ are performed instantaneously after time intervals
of duration $\Delta t$ during which unknown inter-qubit couplings perturb the quantum algorithms 
in the $\tau$-th interaction the ideal quantum gate $U_j$ is replaced by the perturbed unitary quantum gate

\begin{equation}
U_j \mapsto  U_\delta(\tau)  = \exp( - i \delta\!H_{j}^\tau ) U_j 
\label{replace}
\end{equation}
with $\delta H_{j}^\tau  = H_0\Delta t/\hbar$. 
In the notation of (\ref{replace})
the index $\tau = 1,...,t$ takes into account that the perturbations could be different for successive iterations of the algorithm.
However, in the case of static imperfections considered here this possible complication is not relevant.
In Appendix A it is shown that after $t$ iterations and for sufficiently short times $T = t n_g \Delta t$ according to
second order time-dependent perturbation theory
the fidelity amplitude $A$ is given by 
\begin{equation}\label{eq::fstat}
\begin{split}
F_e(t) &= \vert A(t) \vert^2 \\
&= 1 - t \sum_{j,k=1}^{n_g} \frac{1}{N} \tr \{
U_{1\dots j}^\dagger  \delta\!H  U_{j\dots 1} \cdot
U_{1\dots k}^\dagger  \delta\!H  U_{k\dots 1} \} \\
&\hphantom{\mathrel{=}} -2 \sum_{\tau=1}^{t-1} (t-\tau) \sum_{j,k=1}^{n_g} \frac{1}{N} \tr \{
U^{-\tau} \cdot U_{1\dots j}^\dagger  \delta\!H  U_{j\dots 1} \cdot \\
&\hphantom{\mathrel{=}} \qquad\qquad U^\tau \cdot U_{1\dots k}^\dagger  \delta\!H  U_{k\dots 1} \} + \mathcal{O}(\delta\!H^3).
\end{split}
\end{equation}
Thereby, the first term in the sum of (\ref{eq::fstat})
describes the influence of perturbations occurring in the same iteration $\tau$ and the second double sum
describes their influence in different iterations.
This particular form of the short-time behavior of the fidelity amplitude has been studied in detail
by \citeauthor{shep144} \cite{shep144}. In particular, these authors demonstrated that
whenever 
an ideal unitary transformation of a quantum map $U$ can be modeled by a random matrix for sufficiently short times
after $t$ iterations
the corresponding decay of the entanglement fidelity is given
\begin{equation}\label{eq::frahm}
F_e(t) \approx 1 - \frac{t}{t_c} - \frac{2}{\sigma} \frac{t^2}{t_c N} 
\end{equation}
where $1/t_c \approx n_g^2 \tr \{ H_0^2 \} \Delta t^2 /N$ and $\sigma$ denotes the relative fraction of the chaotic
component of the phase space of this map.
Furthermore,
numerical studies indicate that the behavior of higher order terms is such that the fidelity decay becomes approximately  exponential, i.\,e.
\begin{equation}
F_e(t) \approx \exp \Bigl( - \frac{t}{t_c} - \frac{2}{\sigma} \frac{t^2}{t_c N} \Bigr).
\end{equation}

\subsection{Suppression of static imperfections by increasing the decay of correlations}\label{sec::prosen}

In special cases in which an ideal unitary transformation $U$ is not decomposed into elementary gates
we may simplify (\ref{eq::fstat}) by taking $n_g=1$ thus obtaining the short-time fidelity decay
\begin{multline}\label{eq::Prosen1}
F_e(t) =  1 -\\
\sum_{\tau=-(t-1)}^{t-1} ( t - |\tau| ) \frac{1}{N} \tr \{ U^{-\tau} \delta\!H U^\tau \delta\!H \} + \mathcal{O}(\delta\!H^3).
\end{multline}
This expression has been studied previously by \citeauthor{Prosen02} \cite{Prosen02}.
It indicates that the faster the decay of the correlation function
\begin{equation*}
 \frac{1}{N} \tr \{ U^{-\tau} \delta\!H U^\tau \delta\!H \}
\end{equation*}
the slower the decay of the fidelity.
According to an original proposal by \citeauthor{ProZni01} \cite{ProZni01} this characteristic feature of the fidelity decay can be exploited for stabilizing a quantum algorithm against static imperfections.
This aspect was investigated in detail by these authors for the special case of $t=1$.
In this case (\ref{eq::fstat}) reduces to the simpler form
\begin{align}
F_e &=  1 - \sum_{j,k=1}^{n_g} \frac{1}{N} \tr \{
U_{1\dots j}^\dagger  \delta\!H  U_{j\dots 1} \cdot U_{1\dots k}^\dagger  \delta\!H  U_{k\dots 1}
\} + \mathcal{O}(\delta\!H^3) 
\nonumber\\
&\equiv 1 - \sum_{j,k=1}^{n_g}  C(j,k) + \mathcal{O}(\delta\!H^3).
\end{align}
\citeauthor{ProZni01} based their error suppression method on the idea to rewrite a quantum algorithm $U$ in such a way that for the new gate decomposition the sum over the off-diagonal elements of the correlation matrix $C(j,k)$ becomes smaller than for the original gate sequence (thereby using possibly even a larger number of quantum gates).
They considered as an example perturbations of the form $\delta\!H = V \delta$ with $V$ being represented by a $N$-dimensional matrix randomly chosen from the Gaussian unitary ensemble (GUE).
Thus, on average the matrix elements of $V$ fulfill
the condition $\langle V_{jk} V_{lm} \rangle = \delta_{jm} \delta_{kl} / N$.
With this kind of imperfections on average the correlation function becomes
\begin{equation}\label{eq::cfkt}
\langle C(j,k)\rangle =
\delta^2\left(\left \vert \frac{1}{N} \tr \{ U_{j\dots 1} U^\dagger_{1\dots k} \} \right \vert^2 -\frac{1}{N^2}\right).
\end{equation}
The $1/N^2$-term comes from the fact that according to our assumption of traceless perturbing Hamiltonians
also our matrices $V$ have to be chosen traceless.
(In the case of a non-traceless perturbations $V$ this restriction can be achieved by the replacement
$V \mapsto V-\mathcal{I} \tr V /N$).
It should be mentioned that this latter $1/N^2$-term was not taken into account in Ref. \cite{ProZni01} so that these authors investigated the quantity
$\delta^2\left \vert \frac{1}{N} \tr \{ U_{j\dots 1} U^\dagger_{1\dots k} \} \right \vert^2$.

\begin{figure*}
\centerline{\includegraphics[width=\columnwidth]{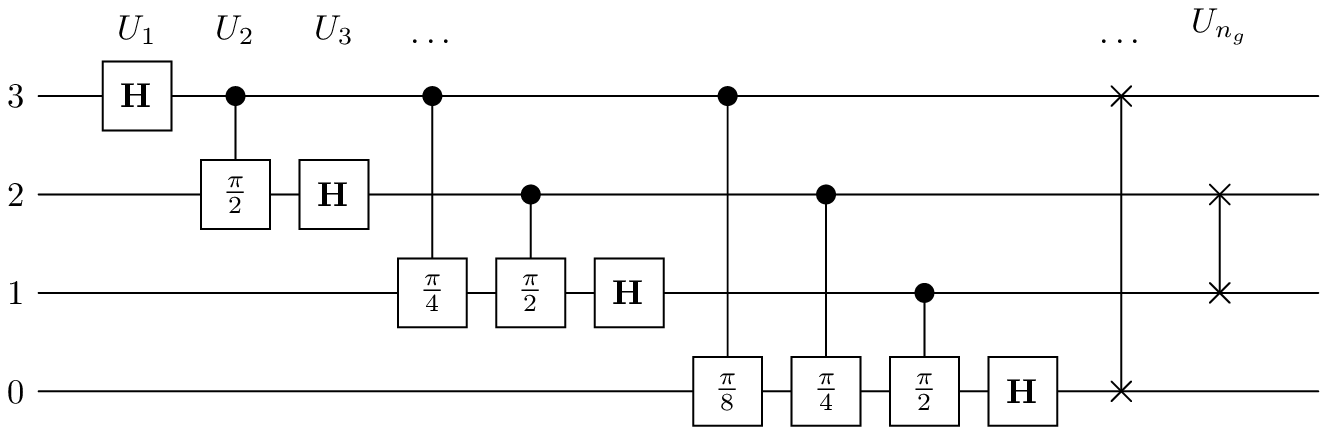}\hfill
\includegraphics[width=\columnwidth]{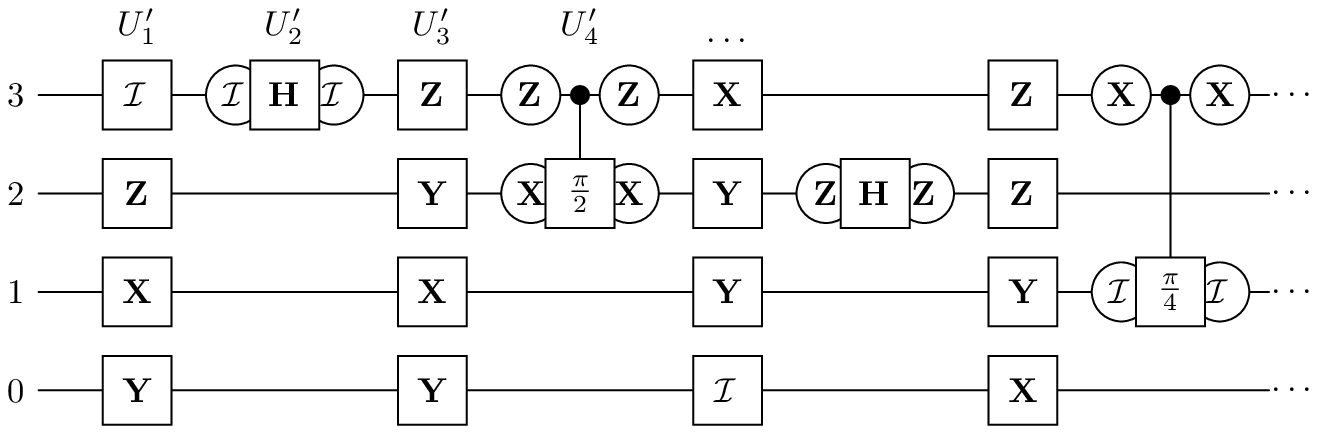}
}
\caption{Quantum circuit of the Quantum-Fourier-Transformation for
$n_q=4$ qubits \textit{(left)}.  The first four gates of the same circuit
involving the PAREC method \textit{(right)}}.\label{fig::qft}
\end{figure*}
\begin{figure*}
\begin{minipage}[b]{0.21\textwidth}
\includegraphics[scale=1.32, trim = 0 41 0 0,clip=true]{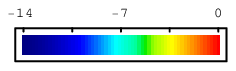}\\
\includegraphics[scale=1.3]{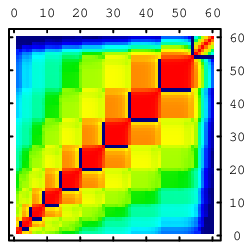}%
\end{minipage}%
\begin{minipage}[b]{0.79\textwidth}
\hfill\includegraphics[scale=1.3]{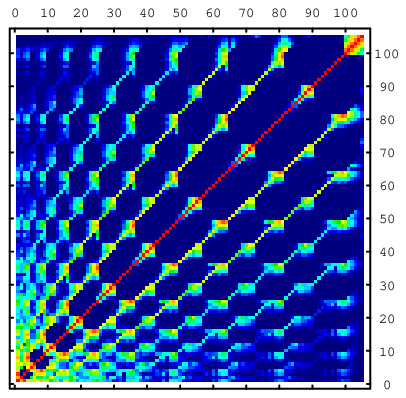}%
\includegraphics[scale=1.3]{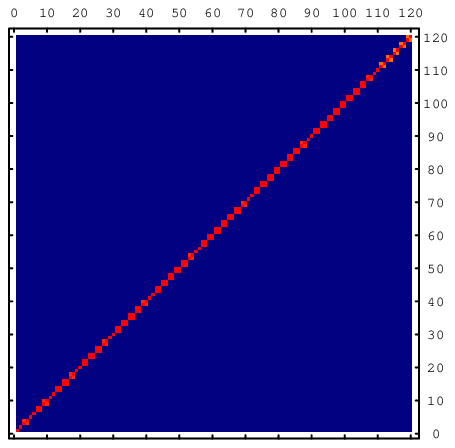}
\end{minipage}
\caption{$\ln \left[( \langle C(j,k) \rangle + 1/N^2)/\delta^2\right] $ for the QFT with $n_q=10$ qubits using the usual gate decomposition with $n_g=60$ gates \textit{(left)},
the decomposition by Prosen using $n_g'=105$ gates \textit{(middle)} and
$\ln \left[( \mathbf{E} \langle C(j,k) \rangle +1/N^2 )/\delta^2\right]$  using the PAREC method with $n_g'=2n_g=120$ gates \textit{(right)}.\label{fig::prosen}}
\end{figure*}

In order to demonstrate their idea, \citeauthor{ProZni01} considered the Quantum-Fourier-Transformation (QFT) as an example.
Typically, this unitary transformation $U$ is decomposed into $n_g = \lfloor n_q(n_q+2)/2 \rfloor$ quantum gates
which involve Hadamard operations and one-qubit phase gates
(compare with the left-hand side of figure \ref{fig::qft}).
Instead, \citeauthor{ProZni01} used a different decomposition involving 
$n_g'=\lfloor n_q(2n_q+1)/2 \rfloor$ quantum gates.
In figure \ref{fig::prosen} the correlation matrix $\langle C(j,k) \rangle$ is depicted for both gate decompositions.
Compared to the conventional gate decomposition (left) the off-diagonal elements of this correlation matrix are suppressed significantly by this new gate decomposition (middle).
Diagonal values are always constant, i.e. $(\langle C(j,j)\rangle + 1/N^2)/\delta^2 = 1$.

Though of interest this proposal of \citeauthor{ProZni01} leaves important questions unanswered. 
How can such an improved gate sequence be found for an arbitrary quantum algorithm ?
How can this be achieved for repeated iterations of a unitary quantum map ?
Is it possible to suppress all off-diagonal elements of the correlation function perfectly ?
As explained in the next subsection all these questions can be addressed and solved in a rather straightforward way utilizing \mbox{\textsf{NRD}} decoupling \cite{parec}.

\subsection{Stabilization of quantum computations by \mbox{\textsf{NRD}} decoupling}\label{sec::parec}
In this section it is demonstrated that the recently proposed Pauli-random-error-correction (PAREC) method \cite{parec} can be understood as a stabilization of a quantum computation using \mbox{\textsf{NRD}} decoupling.
This method allows to stabilize quantum algorithms against static imperfections by converting the typically quadratic time dependences of the resulting fidelity decays \eqref{eq::frahm} into a linear ones.
So far only numerical evidence has been presented for this remarkable property \cite{parec}.
In this section a rigorous formula for the time dependence of the average entanglement fidelity of a PAREC-stabilized iterated quantum algorithm is presented.
It exhibits explicitly the quantitative dependence of this fidelity decay on all relevant characteristic physical parameters.

In the PAREC method before each elementary
quantum gate $U_i,~i=1,...,n_g$ of the $\tau$-th iteration ($\tau=1,...,t$)  of a unitary transformation $U$
a unitary quantum gate 
$r^{(\tau)}_i r^{(\tau)}_{i-1}{}^\dagger$
is applied.
Thereby, the unitary gates $r^{(\tau)}_i$ (with $r^{(1)}_0 = \idty$) are drawn at random from a decoupling set $\mathcal{G}$.
In this particular \mbox{\textsf{NRD}} method (compare with section \ref{sec::random}) $\mathcal{G}$ can always chosen to be an annihilator, such as $\mathcal{P}$ (compare with \eqref{eq::pauli}).
Simultaneously the changes on the quantum algorithm due to these
random unitary gates have to be compensated by replacing each elementary quantum gate $U_i$ of the $\tau$-th iteration of the original algorithm by
$\tilde{U}^{(\tau)}_i = r^{(\tau)}_i U_i r^{(\tau)\dagger}_i$.
Furthermore, after the last quantum gate $\tilde{U}^{(t)}_{n_g}$ a final unitary gate $r^{(t)\dagger}_{n_g}$ is applied.
As a result each iteration of a unitary transformation $U$ is replaced by $2n_g$ unitary quantum gates so that after $t$ iterations one obtains
the result
\begin{equation}\label{eq::uparec}
\begin{split}
&U^t = U  \dots U U = \\
&r^{(t)\dagger}_{n_g} (\tilde{U}^{(t)}_{n_g} \dots r^{(t)}_3r^{(t)\dagger}_2
\cdot \tilde{U}^{(t)}_2 \cdot r^{(t)}_2r^{(t)\dagger}_1{} \cdot \tilde{U}^{(t)}_1 \cdot r^{(t)}_1 r^{(t-1)^\dagger}_{n_g} ) \dots \\
&\qquad
( \tilde{U}^{(2)}_{n_g} \dots r^{(2)}_3r^{(2)}_2{}^\dagger \cdot \tilde{U}^{(2)}_2 \cdot r^{(2)}_2r^{(2)}_1{}^\dagger \cdot \tilde{U}^{(2)}_1 \cdot r^{(2)}_1 r^{(1)}_{n_g}{}^\dagger ) \cdot\\
&\qquad\qquad
( \tilde{U}^{(1)}_{n_g} \dots r^{(1)}_3r^{(1)}_2{}^\dagger \cdot \tilde{U}^{(1)}_2 \cdot r^{(1)}_2r^{(1)}_1{}^\dagger \cdot \tilde{U}^{(1)}_1 \cdot r^{(1)}_1 ) \\
&\equiv (U'^{(t)}_{2n_g} \dots U'^{(t)}_2 U'^{(t)}_1) \dots \\
&\qquad \qquad
        (U'^{(2)}_{2n_g} \dots U'^{(2)}_2 U'^{(2)}_1)
        (U'^{(1)}_{2n_g} \dots U'^{(1)}_2 U'^{(1)}_1)
\end{split}
\end{equation}
with $U^{\prime}_{2k} = \tilde{U}_k,~U^{\prime}_{2k-1} = r^{(\tau)}_k r^{(\tau)}_{k-1}{}^\dagger,~k=1,...,n_g$.
A particular PAREC implementation of the quantum Fourier transform (QFT) is schematically represented
in figure \ref{fig::qft} for the special case of
$n_q=4$ qubits.
Definitely, this random application of Pauli operations together with the associated change of elementary quantum gates does not affect any quantum algorithm.

In order to discuss the stabilizing properties of this particular NRD methods let us consider the static imperfections of the previous section. 
Thus, it is
assumed that each quantum gate $U_i,~i=1,...,n_g$ is performed instantaneously
while between any two of these quantum gates there is a perturbing time evolution originating from the static imperfections
which can be described by the unitary time evolution operator
$\exp ( -i \delta\!H)$ with $\delta\!H = H_0\Delta t/\hbar$.
As shown in Appendix A for sufficiently small times such a static imperfection results
in an average decay of the entanglement fidelity of the form
\begin{align}\label{eq::fidparec}
\mathbf{E} F_e(t) &= 1 - 2 t n_g \frac{1}{N} \tr \{ ( \delta\!H )^2 \} \nonumber\\
& - 2 t \sum_{j=1}^{n_g} \frac{1}{N} \mathbf{E} \tr \{ U^\dagger_j \
r^\dagger \delta\!H r \ U_j \
r^\dagger \delta\!H r \} + \mathcal{O}(\delta^3) \nonumber\\
 &\geq  1 - 4 t n_g \frac{1}{N} \tr \{ ( \delta\!H )^2 \} + \mathcal{O}(\delta\!H^3).
\end{align}
Thereby,
the last inequality can be obtained by recalling that $\tr \{ A^\dagger B \}$ constitutes
a Hermitian inner product for which the Cauchy-Schwarz inequality applies.
Eq. \eqref{eq::fidparec} explicitly exhibits the dependence of the entanglement fidelity decay on the number $n_g$ of elementary quantum gates and
the strictly linear dependence on the numbers of iterations of the unitary transformation $U$.

Several straightforward improvements of the basic relation \eqref{eq::fidparec} are possible.
If the durations $\Delta t_j$ of the static imperfections between adjacent elementary quantum gates are not equal but depend on these gates, for example, \eqref{eq::fidparec} is modified according to
\begin{equation}
\mathbf{E} F_e(t) \geq 1
 - t \frac{1}{N} \tr \{ H_0^2 \}  \sum_{j=1}^{n_g} \bigl( \Delta t_j + \Delta t_{\mbox{bb}} \bigr)^2
+ \mathcal{O}(\delta^3).
\end{equation}
Thereby, $t_{\mbox{bb}}$ denotes the duration of the perturbation after the random pulses.

It is also possible to apply the random gates not before each elementary quantum gate but less often.
One random quantum gate between each iteration of a quantum algorithm, for example, is already
enough to get rid of the terms of \eqref{eq::fstat} quadratic in $t$.
In this case (\ref{eq::fidparec}) is replaced by the inequality
\begin{equation}
\mathbf{E} F_e(t) \geq 1 - t (n_g+1)^2 \frac{1}{N} \tr \{ ( \delta\!H )^2 \}
\end{equation}
at the expense that the term linear in $t$ has a coefficient quadratic in the number of elementary quantum gates per iteration $n_g$.

In order to determine the decay of the average entanglement fidelity of a quantum memory stabilized by \mbox{\textsf{NRD}} decoupling we use (\ref{eq::fidparec}) and specialize to
the case of one iteration $t=1$ of a quantum algorithm consisting of $n_g=n$ identity gates with a time $\Delta t/2$ between subsequent quantum gates.
This way the time between pulses becomes equal to $\Delta t$. Denoting the total interaction time between the qubits of the quantum memory by $T = n_g \Delta t$ one obtains the result
\begin{equation}
\begin{split}
F_e (T=n\Delta t) &\approx 1 - 4 n_g \frac{1}{N} \tr \{ H_0^2 \} (\Delta t/2)^2 \\
 & = 1 - T \Delta t \frac{1}{N} \tr \{ H_0^2 \}.
\end{split}
\end{equation}

\subsection{Static imperfections modeled by the Gaussian unitary ensemble}
In this section it is explicitly shown that the PAREC method is capable of canceling the off diagonal terms of the
correlation function $\langle C(j,k)\rangle$ \eqref{eq::cfkt} perfectly.
Let us consider the stabilization properties of the PAREC method with respect to static imperfections which can be characterized by traceless perturbing Hamiltonians of the form
$\delta H = H_0\Delta t/\hbar \equiv \left(V - \mathcal{I} \tr V / N \right)/\delta $
with $V$ chosen randomly from the Gaussian unitary ensemble (GUE).
These perturbations describe physical situations in which in each individual realization of a quantum
algorithm the inter-qubit Hamiltonian perturbing the dynamics of the qubits of the quantum information processor  is time independent but random.
To eliminate such GUE-governed static imperfections we have to choose an annihilator, such as $\mathcal{P}$,
as a random decoupling set.
As a result the fidelity averaged over all possible random gate reduces to the expression
\begin{equation}
\mathbf{E} \langle F_e \rangle = 1 - \sum_{j,k=1}^{n_g'} \mathbf{E} \langle C(j,k) \rangle + \mathcal{O}(\delta\!H^3).
\end{equation}
According to Eq. \eqref{eq::uparec} in the evaluation of expectation value of the correlation function 
we have to consider $n_g'=2n_g$ quantum gates.
In view of the statistical independence of subsequent Pauli operations
almost all off-diagonal terms of the correlation function vanish, i.e.
{\small
\begin{eqnarray}\label{step}
\frac{ \mathbf{E} \langle C(j,k) \rangle}{\delta^2}  = \begin{cases}
1-\frac{1}{N^2}  &\text{if $j=k$},\\
\vert \frac{1}{N} \tr U_j \vert^2 -\frac{1}{N^2} & \text{if $j$ even and $j=k+1$},\\
\vert \frac{1}{N} \tr U_k \vert^2 -\frac{1}{N^2} & \text{if $k$ even and $k=j+1$}.\\
0 & \text{else}
\end{cases}
\end{eqnarray}}
Thereby, it has been taken into account that for 
all unitary matrices $U$ the relation
\begin{equation}
\mathbf{E} \vert \tr \{ U r \} \vert^2 =
\frac{1}{N^2} \sum_{j=1}^{N^2} \vert \tr \{ U r_j \} \vert^2 = 1
\end{equation}
holds
since the average is performed over all unitary random Pauli gates  $r\in \mathcal{P}$ which are elements of an orthonormal unitary error basis.
As a result the expectation value of the entanglement fidelity becomes
\begin{align}
\mathbf{E} \langle F_e \rangle &= 1 - 2 n_g \delta^2 (1-N^{-2}) \nonumber\\
&\qquad
-2\delta^2 \sum_{j=1}^{n_g} \Bigl( \left\vert \frac{1}{N} \tr U_j \right\vert^2 -N^{-2} \Bigr)
+ \mathcal{O}(\delta^3) \nonumber\\
&\geq 1 - 4 n_g \delta^2 (1-N^{-2}) + \mathcal{O}(\delta^3).
\end{align}
Alternatively this expression can also be derived by substituting the relevant perturbing Hamiltonian $\delta\!H$ 
into \eqref{eq::fidparec}
and setting $t=1$ in \eqref{eq::fidparec}.
For the special case of a quantum Fourier transform (QFT)
the resulting values of $\mathbf{E} \langle C(j,k) \rangle$
are shown on the right-hand side of figure \ref{fig::prosen}.
In this figure they are also compared to the corresponding values resulting from the improved QFT proposed by Prosen.


\section{Combining dynamical decoupling and detected-jump error correction}
In this section a combined error correction scheme is introduced which allows a restricted form of dynamic decoupling within a detected-jump error correcting quantum code space.
The error suppressing properties of this combined error correcting scheme are investigated in the presence of spontaneous
decay processes and coherent but unknown Heisenberg-type inter-qubit couplings within a quantum information processor.
Simple approximate expressions are derived for the expected average fidelity 
decay of instantaneous decoupling sequences, which are then confirmed by 
numerical simulations.

\subsection{Spontaneous decay}

\subsubsection{Fidelity decay due to spontaneous decay}
If the qubits of a quantum information processor  couple to their environment,
spontaneous state changes can occur by emitting energy in form of photons, for example.
Such processes cause state changes from an excited state of a qubit, say $\ket{1}$, to an energetically lower lying state, say $\ket{0}$.
Let us assume for our subsequent discussion that this state $\ket{0}$ is the qubit's stable ground state.
The dynamics of spontaneous decay processes based on photon emission  can be described by a master equation in Lindblad form \cite{Carm99}
\begin{equation}\label{masterequation}
\dot{\rho} = -\frac{i}{\hbar} [H_{S}(t), \rho] + \frac{1}{2} \sum_{i=0}^{n_q-1} ([L_i, \rho L_i^\dagger] + [L_i \rho, L_i^\dagger]).
\end{equation}
Here, $\rho$ describes the density operator of the quantum register involving $n_q$ distinguishable qubits.
Their unitary dynamics are described by the Hamiltonian
$H_S(t)$. Typically, these dynamics originate from externally applied quantum gates.
The non-unitary aspects of the time evolution of the quantum register are described by the Lindblad operators
\begin{equation}
L_k = \sqrt{\kappa_k} \ket{0}_{i\,i}\bra{1} \bigotimes \idty_{j\neq k},~~~k=1,...,n_q
\end{equation}
which describe
the influence of spontaneous emission processes on the qubits of the quantum register.
In \eqref{masterequation} it is assumed that the distinguishable qubits are separated by a distance
much larger than the wavelength of the spontaneously emitted photons. As a result, to a good degree of approximation
the qubits couple to statistically independent reservoirs \cite{Carm99}. If the decay rates of the qubits $\kappa_k$ are all equal,
it is possible to construct quantum codes capable of correcting the spontaneous-emission induced errors.

However, before addressing such error correcting quantum codes let us look briefly at the expected fidelity decay
of a quantum memory under the influence of such spontaneous decay processes.
For this purpose let us consider the fidelity
\begin{equation}
F(t) = \langle \psi(0) \vert \rho(t) \vert \psi(0)\rangle,
\end{equation}
with $\rho(t)$ obeying Eq.(\eqref{masterequation}) with the initial condition $\rho(0)=\ket{\psi(0)}\bra{\psi(0)}$.
If a spontaneous decay event affects qubit $i$ an arbitrary state of of this qubit is mapped onto its ground state
$\ket{0}_i$. As a consequence the resulting $n_q$-qubit state is expected to have almost no overlap with the  $n_q$-qubit state
before the decay process. Therefore, the fidelity is expected to become very small as soon as an decay process has affected any of the
$n_q$ qubits.
If there are $n_1$ qubits in excited states initially,
the probability that no decay event takes place in a time interval of duration $t$ is
given by
\begin{equation}\label{fidelity non-corrected decay}
p(t) = \exp(-n_1 \cdot \kappa \cdot t)
\end{equation}
provided the 
spontaneous decay rate $\kappa$ is the same for all qubits.
Thus.  assuming that the fidelity drops to zero as soon a spontaneous decay process has taken place
the average expected fidelity at time $t$ is approximately given by $F(t) \approx p(t)$ \cite[see also][]{kernjmp}.

\subsubsection{Detected-Jump Correcting Codes}
Spontaneous decay events as described by \eqref{masterequation} can be corrected by
detected-jump correcting quantum codes \cite{ABC01}.
Thereby, all modifications of the ideal dynamics which do not arise from any photon decay events
are corrected passively by an encoding of the logical information in an appropriate decoherence free subspace (DFS).
This DFS is spanned by all quantum state involving a constant number of excited (physical) qubits. In order to maximize the dimension of
this DFS one may choose quantum state with $n_q/2$ excited qubits.
In order to be able to correct also any single error due to spontaneous decay of a known qubit by unitary recovery operations 
one has to construct code words by bitwise complementary pairing. 
In the case of four physical qubits, for example, such a one-error detected-jump correcting quantum code space is spanned by the
three orthogonal logical states
\begin{subequations}
\begin{align}
\ket{c_0} &= \frac{1}{\sqrt{2}} (\ket{0011} + e^{i\varphi}\ket{1100}),\\
\ket{c_1} &= \frac{1}{\sqrt{2}} (\ket{0101} + e^{i\varphi}\ket{1010}),\\
\ket{c_2} &= \frac{1}{\sqrt{2}} (\ket{0110} + e^{i\varphi}\ket{1001}).
\end{align}
\end{subequations}
Thereby, the value of relative phase $\varphi$ can be chosen arbitrarily. A detailed discussion of these error correcting quantum codes and
their recovery operations can be found in Ref. \cite{alber2}. In particular, it has been shown that 
provided the error position and error time is known the recovery operation
\begin{equation}\label{recoveryoperation}
R_i = X_i \cdot \prod_{j \neq i} \CNOT_{i, j} \cdot H_i
\end{equation}
restores any quantum state perfectly.
The time $t_{rec}$ needed for this recovery operation can be estimated by the time needed for the performance of $n_q-1$ controlled NOT gates.

In order to simplify computation on the code space it is advantageous to use subspaces of one-error detected-jump correcting quantum codes
which are equipped with a formal tensor product structure and have been introduced by \cite{khod1}.
The main idea is to map
each logical qubit onto two physical qubits, i.e.  $\ket{0}_L \rightarrow \ket{01}_P$ and $\ket{1}_L \rightarrow \ket{10}_P$, and 
again supplementing these code words with their bitwise complements.
In order to be able to distinguish between logical two-qubit code words, such as $\ket{00}$ and $\ket{11}_L$, for example,
on has to introduce two additional physical ancilla qubits. Thus, for two logical qubits, for example, the resulting four code words are given by
\begin{subequations}
\begin{align}
\ket{00}_L &\rightarrow \frac{1}{\sqrt{2}} (\ket{01~01~01} + \ket{10~10~10}),\\
\ket{01}_L &\rightarrow \frac{1}{\sqrt{2}} (\ket{01~10~01} + \ket{10~01~10}),\\
\ket{10}_L &\rightarrow \frac{1}{\sqrt{2}} (\ket{10~01~01} + \ket{01~10~10}),\\
\ket{11}_L &\rightarrow \frac{1}{\sqrt{2}} (\ket{10~10~01} + \ket{01~01~10}).
\end{align}
\end{subequations}
This scheme encodes $n_L$ logical qubits in $n_P = 2\,n_L + 2$ physical qubits. As outlined in \cite{khod1} and \cite{kernjmp}
it is possible to construct simple universal logical gates on these
code spaces. 
With these quantum codes any single spontaneous decay process can be corrected
perfectly provided the appropriate recovery operation is performed perfectly.

Let us look briefly at the expected average fidelity decay of a quantum memory which is protected by these jump codes and whose qubits decay spontaneously.
The probability that no spontaneous decay takes place during a recovery process is given by 
\begin{equation}
p^\prime (t_{rec}) = \exp(-n_q/2 \cdot \kappa \cdot t_{rec})
\end{equation}
since on average at most $n_q/2$ qubits are excited during the recovery operation.
Up to a time $t$ one expects an average number $n_e(t) = \frac{n_q}{2} \kappa t$ of decay events. 
Thus, at time $t$ the fidelity of the quantum memory is approximately given by \cite{kernjmp}
\begin{equation}\label{fidelity jumpcodes}
F(t)= p^\prime(t_{rec})^{n_e(t)} = \exp(- \left(\frac{n_q \kappa}{2}\right)^2 t_{rec} t).
\end{equation}
Comparing (\ref{fidelity jumpcodes})
with the corresponding uncorrected fidelity decay of (\ref{fidelity non-corrected decay}) (with $n_1 = n_q/2$)
one notices a significant improvement provided
the number of qubits $n_q$ is too large, i.e. $\kappa n_q t_{rec} /2 \ll 1$.
However, for large numbers of qubits, i.e. $\kappa n_q t_{rec} /2 > 1$, this method does not work.
This shortcoming can be remedied by encoding the physical qubits in blocks so that each
block can
correct one decay event at a time and by using block-entangling gates, such as the ones introduced in Ref. \cite{AMD02, kernjmp}.
As a result error correction 
fails only if more than one spontaneous decay process occurs in the same block during recovery and the fidelity decay can be suppressed considerably
in comparison with (\ref{fidelity jumpcodes}).

\subsection{Embedding EMD decoupling inside a one-error detected-jump correcting quantum code space \label{coherentcouplings}}

Let us now investigate to which extent additional errors originating from uncontrolled inter-qubit couplings can be corrected by embedding
a suitably chosen decoupling procedure into the code space of a one-error detected-jump correcting quantum code.

For this purpose let us consider a case in which in addition to spontaneous decay events  also static energy detunings and nearest neighbor interactions
affect the physical qubits of a quantum memory.
Correspondingly, the error Hamiltonian describing these inter-qubit couplings is given by
\begin{equation}\label{coherenthamiltonian}
\begin{split}
H_S(t) &= \sum_{i=0}^{n_q-1} \hbar\delta_i Z_i + \sum_{K = X, Y, Z} \sum_{j=0}^{n_q-2} \hbar J_{K,j} K_j K_{j+1}\\
       &\equiv H_0,
\end{split}
\end{equation}
with
$X$, $Y$ and $Z$ denoting the Pauli spin operators. Thereby, the quantities $\delta_i$
denote detunings which may arise from an external magnetic field, for example, and the parameters $J_{K,j}$ characterize the strengths
of nearest neighbor couplings. In \eqref{coherenthamiltonian} it is
assumed that the qubits are arranged on a linear chain.

The detected-jump correcting quantum codes introduced above rely on a DFS involving a constant number of excited qubits.
Therefore, one cannot use arbitrary decoupling schemes for suppressing the 
influence of unwanted inter-qubit couplings, since in general they may also change the number of excitations
in the code words and as a result the code space would be left.
Within the set of Pauli operations  only the $Z$ and swap-type operations
leave these DFSs invariant. So, we can use a combination of $Z$-based \emph{flips} and a random decoupling based on the symmetric group
$\mathcal{S}_{n_q}$ of permutations acting on $n_q$ qubits (\emph{swaps}) for performing dynamical decoupling inside a
one-error detected-jump correcting quantum code space.
Thus, according to the notions of section \ref{sec::DD} this constitutes an \mbox{\textsf{EMD}} decoupling method,
embedding \mbox{\textsf{PDD}} flip-decoupling inside \mbox{\textsf{NRD}} swap-decoupling.
In particular,
flip-decoupling applies the flip-operation
\begin{equation}
U_F = Z \otimes \idty \otimes Z \otimes \ldots \otimes \idty
\label{flipop}
\end{equation}
after time intervals of duration, say $\tau$. This has the effect of canceling all 2-qubit couplings of $XX$- and $YY$-type and cross terms involving
$X$ and $Y$ Pauli operations up to second order average Hamiltonian theory. But it leaves $Z$- and $ZZ$-type couplings invariant.
Thus,
neglecting higher order terms caused by $XX$- and $YY$-type couplings the resulting intermediate average Hamiltonian $H^\prime_0$ is given by
\begin{equation}
H^\prime_0 = \sum_{i=0}^{n_q-1} \hbar \delta_i Z_i + \sum_{j=0}^{n_q-2} \hbar J_{z,j} Z_j Z_{j+1}.
\end{equation}
It should be noted that this type of \mbox{\textsf{EMD}}
decoupling method is applicable only if the number
of physical qubits used in the code is a multiple of four with a corresponding odd number of logical qubits.
Otherwise the flip operation (\ref{flipop}) maps some codewords from symmetric ($\varphi =0$) to antisymmetric ($\varphi =\pi$)
superposition states which causes the
recovery to fail.
After a time $\Delta t = m\, \tau$, where $m$ is an even integer, swap-decoupling permutes the logical qubits 
by executing up to $n_q-1$ pairwise swap operations (transpositions). These swap operations
are chosen at random in such a way that all permutations of the logical qubits  are equally probable.
Keeping track of the applied permutations, one knows which physical qubit one has to address in order to access the information
that was originally stored in, say, the first qubit.
As a result of this procedure
decoupling results from the fact that the code words effectively see changing coupling constants between their qubits.

\subsubsection{Expression for the expected fidelity of flipswap decoupling}
To estimate the fidelity decay of a quantum memory perturbed by $H_0$ and protected by the flipswap decoupling discussed in Sec.\ref{coherentcouplings},
we calculate the evolution of a pure state, say $\ket{\psi}$, under the influence of the effective toggling-frame error Hamiltonian, i.e.
\begin{equation}\label{errorhamiltonian}
U(t = N \Delta t) = \mathcal{T} \prod_{j=0}^{N} \exp(-i H^{(j)}_0 \Delta t/\hbar)
\end{equation}
with
\begin{equation}
H^{(j)}_0 = P^\dagger_j H_0^\prime P_j
\end{equation}
and $P_j$ denoting the toggling-frame Hamiltonian 
and the  permutations chosen randomly and independently for each time interval.
Up to second order perturbation theory
after time $t$ the quantum state is given
by
\begin{multline}\label{secondorderlong}
\ket{\psi(t)} = \Bigl( \idty -i \sum_{j=1}^N  H^{(j)}_0 \Delta t/\hbar  \\
- \sum_{i=2}^{N} \sum_{j=1}^{i-1}  H^{(i)}_0  H^{(j)}_0 ({\Delta t}/\hbar)^2
- \frac{1}{2} \sum_{j=1}^N {H^{(j)}_0}^2 ({\Delta t}/\hbar)^2 \Bigr) \ket{\psi(0)}
\end{multline}
Comparing this state with the ideal state $\ket{\psi}$ yields the fidelity 
\begin{equation}\label{fidelityrandomswap}
\begin{split}
F_C &= \Bigl\langle \abs{ \braket{\psi}{\psi(t)} }^2 \Bigr\rangle_P  \\
&\approx 1- \sum_{i,j=1}^N \Bigl\langle  \bra{\psi} \bigl(  H^{(i)}_0 - \bra{\psi} H^{(i)}_0 \ket{\psi} \bigr) \cdot \\
&\qquad\qquad\qquad\quad \bigl(  H^{(j)}_0 - \bra{\psi} H^{(j)}_0 \ket{\psi} \bigr) \ket{\psi}  \Bigr\rangle_P ({\Delta t}/\hbar)^2 =\\
&=1- N({\Delta t}/\hbar)^2 \Bigl(  \bra{\psi} \Bigl\langle H'_0{}^2 \Bigr\rangle_P \ket{\psi} -
\Bigl\langle \bra{\psi}H'_0\ket{\psi}^2 \Bigr\rangle_P \Bigr)\\
&\, - (N^2-N)({\Delta t}/\hbar)^2 \Bigl( \bra{\psi} \Bigl\langle H'_0\Bigr\rangle_P^2 \ket{\psi} -
 \Bigl\langle \bra{\psi}H'_0\ket{\psi} \Bigr\rangle_P^2 \Bigr)
\end{split}
\end{equation}
with
$\langle \ldots \rangle_P$ denoting the ensemble average over all possible permutations,
\begin{equation}
\langle A \rangle_P = \frac{1}{n_q !} \sum_{P_i \in \mathcal{S}_{n_q}} P_i^\dagger A P_i.
\end{equation}
In order
to find an explicit expression for the expected fidelity
we need to calculate the ensemble averages of $H'_0$, $H'_0{}^2$ and $\bra{\psi}H'_0\ket{\psi}^2$.

Let us start with the calculation of $\langle H'_0 \rangle_P$.
Since the initial state $|\psi\rangle$ is an encoded quantum state
we are only interested in the part of the Hamiltonian which has its support on the code space. Therefore, it is convenient to  
introduce the projection operator on this code space $\Pi$.
By combinatorial arguments
one can show
that the average of the term involving the detunings $\delta_i$ is given by
\begin{equation}
\Pi \Bigl\langle \sum_{i=0}^{n_q-1} \hbar \delta_i Z_i  \Bigr\rangle_P  \Pi = 0.
\end{equation}
This is due to the fact that
each qubit is influenced by any detuning equally often and that inside the code space exactly half of the qubits are in excited states.
The averaging of the terms involving $ZZ$-type interactions is slightly more complicated.
However, on the code space one obtains the relation
\begin{equation}
\begin{split}
\Pi \Bigl\langle  \sum_{i=0}^{n_q-1} \hbar J_{z, i} Z_i Z_{i+1}  \Bigr\rangle_P \Pi
&= -\frac{1}{n_q-1} \sum_{i=0}^{n_q-1} \hbar J_{z,i} \Pi \\
& =: c_1 \Pi
\end{split}
\end{equation}
with
the real number $c_1$ depending on the strengths of the inter-qubit couplings $J_{z,i}$.
Similarly, one can calculate $\langle H'_0{}^2 \rangle_P$.
Cross terms of $Z$- and $ZZ$-type vanish after ensemble averaging.
Remaining terms are proportional to $\idty$, $ZZ$ or $ZZZZ$. Averaging these latter terms over all permutations results in
\begin{equation}
\begin{split}
\Pi \langle H_0^{\prime 2}/\hbar^2 \rangle_P \Pi&= \Bigl[
 \sum_{i=0}^{n_q-1} \delta_i^2 -\frac{2}{n_q-1} \sum_{i<j} \delta_i \delta_j \\
&+\sum_{j=0}^{n_q-2} J_{z,j}^2  -\frac{2}{n_q-1}\sum_{i=0}^{n_q-3} J_{z,i}J_{z,i+1}\\
&\qquad +2 (p_{+}^{\prime}-p_{-}^{\prime}) \sum_{j=2}^{n_q-2} J_{z,i}J_{z,j} \Bigr] \Pi\\
&=: c_2 \Pi /\hbar^2,
\end{split}
\end{equation}
with
$p^\prime_{\pm}$ denoting the probability of finding four bits of a word with $n_q/2$ excited qubits arranged in such a way
that applying the four-qubit operator $ZZZZ$ to them yields the eigenvalue $\pm 1$. Furthermore, one finds the relation 
\begin{equation}
p_{+}^{\prime}-p_{-}^{\prime} = \frac{3}{3-4n_q+n_q^2}.
\end{equation}
The last term one has to calculate is given by
\begin{equation}
\Bigl\langle \bra{\psi}H_0'\ket{\psi}^2 \Bigr\rangle_P =: c_3,
\end{equation}
and depends on the initial state $\ket{\psi}$.
This initial state can always be expressed as a superposition of the symmetric ($\varphi =0$)
basis states, i.e. $\ket{\psi} = \sum_k a_k \ket{\psi_k}$ with $\sum_k |a_k|^2 = 1$.
Therefore,
detunings do not matter, since the application of any $Z_i$ operator to any symmetric basis state changes it into its
antisymmetric pendant and vice versa.
It is not difficult to to prove the following upper and lower bounds for the positive constant $c_3$
\begin{equation}
 c_1^2 \leq c_3 \leq \bigl(c_2 - \sum_{i=0}^{n_q-1} \hbar^2 \delta_i^2 +\frac{2\hbar^2 }{n_q-1} \sum_{i<j} \delta_i \delta_j\bigr).
\end{equation}
In general,
the value of the quantity $c_3$ depends on the quantum state.
The upper bound corresponds to
the case $a_i=\delta_{ii_0}$ and the lower bound to the case $a_i=\text{const}$.

Inserting these results into the fidelity \eqref{fidelityrandomswap}, the qua\-dratic-in-time term vanishes since
$\Pi \langle H_0^\prime \rangle_P \Pi\propto \Pi$.
With $t_0 = 0$ the fidelity due to coherent errors becomes independent of the initial state $|\psi\rangle$ and is given by
\begin{equation}
F_C \approx 1 - t \, \Delta t (c_2-c_1^2)/\hbar^2
\end{equation}
with
$\Delta t$ denoting the time between subsequent swap operations. 
Thus, a linear-in-time fidelity decay is obtained due to the fact that the EMD decoupling suppresses the quadratic-in-time decay
resulting from inter-qubit couplings.  This is a main results of this section.
This calculated linear can be used as the leading order of an exponential approximation yielding the fidelity decay 
\begin{equation}\label{heuristic coherent fidelity}
F_C \approx \exp\bigl(- t \, \Delta t (c_2-c_1^2)/\hbar^2 \bigr).
\end{equation}

\subsubsection{Implementations and Numerical Simulations}
\begin{figure}
\includegraphics[width=\columnwidth]{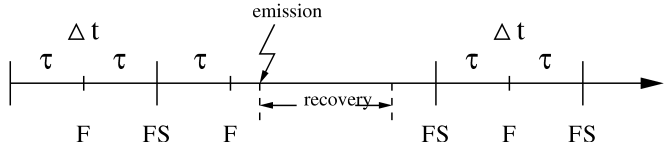}
\caption{A typical sequence of instantaneously applied
flip-type (F) and swap-type (S) decoupling operations and of a recovery operation of duration $t_{rec}$:
During recovery no decoupling operations are executed. Here the time delay between swap operations $\Delta t$
and the time delay between subsequent flip operations $\tau$ are related by $\Delta t = 2\tau$.
\label{instadecoupling}}
\end{figure}

In a practical implementation of the combined scheme discussed above one corrects
spontaneous decay events with priority by taking into account that recovery operations must not be interrupted by decoupling operations.
This compatibility requirement guarantees that no additional uncorrectable errors occur during the recovery steps.
A corresponding typical correction sequence is depicted schematically in
Fig. \ref{instadecoupling}.
This compatibility requirement
may lead to varying time intervals between individual decoupling operations. However, for the decoupling scheme discussed here this does not present
any serious problems. Assuming that
a successful recovery operation recovers a faulty state perfectly while a failed recovery approximately results in a vanishing fidelity,
the fidelity decay of such a combined error correcting scheme is given by products of the fidelities of 
Eqs. \eqref{fidelity jumpcodes} and \eqref{heuristic coherent fidelity}.

In the following we discuss the fidelity decay in the idealized case of instantaneous decoupling operations.
However, the recovery operation is assumed to require always a finite total time which grows linearly with the number of qubits involved (compare with \eqref{recoveryoperation}),
\begin{equation}
 t_{rec} = \bigl( \frac{3}{2} (n_q-1) +2 \bigr)t_0.
\end{equation}
Therefore, secondary spontaneous decays during a recovery cause errors.
Composite gates, such as CNOT gates, are assumed to take $3/2$ times as long as the basic gate time $t_0$.

\subsubsection{Instantaneously Applied Decoupling Operations}
Let us assume
that the deterministic flip operations and stochastic swap operations are performed instantaneously after time intervals of duration $\tau$ and
$\Delta t = m\tau$ with an even integer $m$
(compare with Fig.\ref{instadecoupling}).
Whenever a spontaneous decay event occurs between two flip operations 
the recovery operation is applied and the decoupling sequence is halted for a time interval of duration $t_{rec}$.
Furthermore, 
we are primarily interested in cases of frequent decouplings, i.e. $t_{rec} \gg \tau$,
with spontaneous decay representing a weak perturbation, i.e.
$\kappa \tau\ll 1$.
Thus,
according to Eqs.
\eqref{fidelity jumpcodes} and \eqref{heuristic coherent fidelity} 
after time $t = N\Delta t$
the expected fidelity $F_{ID}$ of a quantum memory whose dynamics are governed by the master equation \eqref{masterequation} with Hamiltonian \eqref{coherenthamiltonian} is approximately given by
\begin{eqnarray}\label{ffinal}
F_{ID}&=&
\exp\Bigl( -(c_2-c_3) (t \Delta t/\hbar^2) \left(1 + \kappa t_{rec}n_q/2 \right)\Bigr) \times \nonumber\\
&& \exp\Bigl( -(\frac{n_q \kappa}{2}\bigr)^2 t_{rec} t \Bigr)
\end{eqnarray}
with $n_{ph} = (n_q/2)\kappa t$ denoting the mean number of spontaneous decay events taking place during the time interval 
$t = N\Delta t$.
Thus, in order to maximize this fidelity
it is optimal to
minimize the decoupling time interval $\Delta t$.

\subsection{Numerical Results\label{numericalsimulations}}
Let us assume that the coupling constants $\delta_j$ and $J_{z,j}$ of the Hamiltonian \eqref{coherenthamiltonian} are time independent but unknown and that they are distributed randomly and uniformly in the interval $[-\sqrt{3} \varepsilon,\sqrt{3} \varepsilon]$.
In the following we discuss the time evolution of the fidelity of a quantum memory with three logical qubits in a coherent state perturbed by spontaneous decay processes and by inter-qubit couplings (compare with Eqs. \eqref{masterequation} and \eqref{coherenthamiltonian}.
The numerical simulations were performed with the help of the quantum trajectory method \cite{dumpazo}.
Random elements which had to be averaged over several trajectories were the realizations of random swap operations on the one hand and times and positions of spontaneous decay processes on the other hand.
\begin{figure}
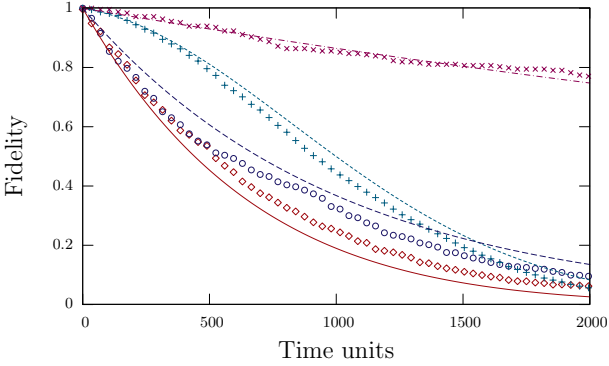

\include{k3e4-insta}
\caption{\label{fig e-4 instantan} Time evolution of the average fidelity whose dynamics are governed by Eq.(\ref{masterequation}):
3 physical qubits without any error correction or decoupling (red $\lozenge$),
3 physical qubits with instantaneous dynamical decoupling only with $\Delta t = 4t_0$ (blue $\circ$),
8 physical qubits encoding 3 logical qubits with combined error correction and instantaneous decoupling with $\Delta t = 4t_0$ (pink $\times$),
8 physical qubits encoding 3 logical qubits without decoupling (light blue $+$).
(The time is plotted in units of the elementary gate time $t_0$.)
The full lines represent the corresponding approximate analytical expressions given in the text.
The parameters are $\kappa t_0 = 10^{-3}$, $\varepsilon t_0= 10^{-4}$.
Statistical fluctuations originate from the sampling of the permutations.
In the curve for $n_q=8$, for example, 500 random permutations are chosen per quantum trajectory over a total of 100 trajectories
which is only slightly more than the size of the permutation group ($8! = 40320$).}
\end{figure}

In the case of instantaneous decoupling operations it is best to apply them as often as possible
as assumed in \eqref{ffinal}, for example.
Fig. \ref{fig e-4 instantan} shows a 3-qubit and an 8-qubit encoded quantum memory.
The curves drawn correspond to the cases of no decoupling and error correction at all (red $\lozenge$),
of decoupling only (blue $\circ$), of error correction by a jump code only (light blue $+$), and of
the combined scheme discussed here (pink $\times$).
The characteristic quadratic-in-time fidelity decay due to coherent errors is apparent from
the curves without decoupling
(blue $\circ$). This decay is
superimposed by a (fast) linear-in-time decay due to (un)corrected spontaneous decay.
The more frequently one applies decoupling operations, the closer one gets to the spontaneous-decay-induced limit described by
\eqref{fidelity jumpcodes}.

\section{Summary}
Two applications for random decoupling have been presented.

In section 3 it was shown that random decoupling can be used to stabilize quantum algorithms against static imperfections
in an efficient way. In particular, a simple expression was derived 
for the average fidelity decay of a quantum memory emerged which complements already known
rigorous lower bounds for the worst case behaviour.

In section 4 it has been demonstrated that errors originating from uncontrolled couplings to an environment and from couplings
among qubits of a quantum information processor
can be corrected efficiently by combining redundancy-based quantum error correction with dynamical decoupling methods.
In the particular example presented an embedded dynamical decoupling scheme was embedded into a one-error correcting detected-jump correcting quantum code.
This way both spontaneous decay processes and uncontrolled inter-qubit couplings could be suppressed significantly.
An important aspect which has to be taken into account in the development of such combined error suppression methods is compatibility.
It has to be ensured that the dynamical decoupling operations used do not leave the code space of the redundancy-based
quantum error correcting quantum code.
Typically this compatibility requirement constrains the structure of the admissible dynamical decoupling operations.

\begin{acknowledgement}
This work is supported by the DAAD.
I.~J. acknowledges also financial support by project MSM 6840770039.
Stimulating discussions with Dima L. Shepelyansky are acknowledged.
\end{acknowledgement}

\appendix

\section{Appendix}
In this appendix the perturbative short-time approximation of the fidelity amplitude used in (\ref{eq::fstat}) is derived.
Let us consider a quantum algorithm iterating an ideal unitary transformation
$U= U_{n_g}\cdots U_3\cdot U_2\cdot U_1$
$t$ times. If this ideal time evolution
is perturbed by static imperfections  originating from Hamiltonian inter-qubit couplings
the $j$-th quantum gate
of the $\tau$-th iteration of $U$
is replaced by
the perturbed unitary quantum gate
\begin{equation}
U_j \mapsto  \exp( - i \delta\!H_{jl}^\tau )
U_j \exp( - i \delta\!H_{jr}^\tau ).
\label{replace1}
\end{equation}
In Eq.(\ref{eq::fstat}), for example, the special case $\delta\!H_{jl}^{\tau} = \delta H = H_0 \Delta t$ and
$\delta\!H_{jr}^{\tau} = 0$ is considered. The index $\tau$ in (\ref{replace1}) takes into account that perturbations may be different
in successive iterations of the unitary transformation $U$.
In a second order expansion
with respect to $\delta\!H_{jl}^\tau$ and $\delta\!H_{jr}^\tau$ 
after $t$ iterations 
the fidelity amplitude $A$ (\ref{fidampl})of the iterated perturbed
quantum algorithm is given by
\begin{equation}\label{eq::secorder}
\begin{split}
A(t) &= 1 - \sum_{p=l,r}\sum_{\tau=1}^t\sum_{j=1}^{n_g} \frac{1}{N}\Bigl(
i\tr \{ \delta\!H_{jp}^\tau \}  - \frac{1}{2}\tr \{ (\delta\!H_{jp}^\tau)^2 \} 
\Bigr)\\
&-\sum_{\tau=1}^t\sum_{j=2}^{n_g}\sum_{k=1}^{j-1}\frac{1}{N}\Bigl[
  \tr \{ \delta\!\tilde{H}_{jl}^\tau(j)  \delta\!\tilde{H}_{kl}^\tau(k) \} \\
& +\tr \{ \delta\!\tilde{H}_{jl}^\tau(j)  \delta\!\tilde{H}_{kr}^\tau(k-1) \}\\
& +\tr \{ \delta\!\tilde{H}_{jr}^\tau(j-1)  \delta\!\tilde{H}_{kl}^\tau(k) \} \\
& +\tr \{ \delta\!\tilde{H}_{jr}^\tau(j-1)  \delta\!\tilde{H}_{kr}^\tau(k-1) \}  \Bigr]\\
&-\sum_{\tau=1}^t\sum_{j=1}^{n_g}\frac{1}{N} \tr \{ U^\dagger_j \delta\!H_{jl}^\tau U_j \delta\!H_{jr}^\tau \}\\
&-\sum_{\tau_1=2}^t\sum_{\tau_2=1}^{\tau_1-1}\sum_{j,k=1}^{n_g}\frac{1}{N}\Bigl[\\
&\mathrel{\phantom{+}}
  \tr \{ U^{\tau_2-\tau_1} \delta\!\tilde{H}_{jl}^{\tau_1}(j) U^{\tau_1-\tau_2}
\delta\!\tilde{H}_{kl}^{\tau_2}(k) \} \\
&+\tr \{ U^{\tau_2-\tau_1} \delta\!\tilde{H}_{jl}^{\tau_1}(j) U^{\tau_1-\tau_2} \delta\!\tilde{H}_{kr}^{\tau_2}(k-1) \} \\
&+\tr \{ U^{\tau_2-\tau_1} \delta\!\tilde{H}_{jr}^{\tau_1}(j-1) U^{\tau_1-\tau_2} \delta\!\tilde{H}_{kl}^{\tau_2}(k) \} \\
&+\tr \{ U^{\tau_2-\tau_1} \delta\!\tilde{H}_{jr}^{\tau_1}(j-1) U^{\tau_1-\tau_2} \delta\!\tilde{H}_{kr}^{\tau_2}(k-1) \}
\Bigr] \\ &+ \mathcal{O}(\delta\!H^3)
\end{split}
\end{equation}
with
\begin{equation}
\begin{split}
\delta\!\tilde{H}_{jp}^\tau (i) &= U_1^\dagger U_2^\dagger \dots U_i^\dagger \cdot \delta\!H_{jp}^\tau \cdot U_i \dots U_2 U_1 \\
&= U_{1\dots i}^\dagger  \, \delta\!H_{jp}^\tau \, U_{i\dots 1}.
\end{split}
\end{equation}
The term linear in the perturbing Hamiltonian $\delta\!H$ vanishes if all Hamiltonians involved are traceless.
Note that all the terms of 
(\ref{eq::secorder})
involving
$\tr \{ \cdot \}$ terms
are real valued so that up to second order the fidelity $F_e(t) = \vert A(t) \vert^2$ 
is simply obtained
by multiplying all these terms of $A(t)$ with a factor of magnitude two. 
In the special case 
$\delta\!H_{jl}^{\tau} = \delta H = H_0 \Delta t$, 
$\delta\!H_{jr}^{\tau} = 0$ (\ref{eq::secorder}) gives rise to Eq.(\ref{eq::fstat}).

Setting $\delta\!H^\tau_{il} = \delta\!H^\tau_{ir} = r^{(\tau)}_i{}^\dagger \delta\!H r^{(\tau)}_i$
Eq. \eqref{eq::secorder} 
yields the
time evolution of \eqref{eq::uparec}.
Performing an average $\mathbf{E}$ over all random unitary operations $\{ r^{(\tau)}_i \}$ from the
decoupling set 
$\mathcal{G}$
one obtains the expectation value  
of the amplitude
\begin{multline}
\mathbf{E} A(t) = 1
 - \sum_{\tau=1}^t \sum_{j=1}^{n_g} \frac{1}{N} \tr \{(
r^{(\tau)}_j{}^\dagger \delta\!H r^{(\tau)}_j
)^2 \} \\
- \sum_{\tau=1}^t \sum_{j=1}^{n_g} \frac{1}{N} \mathbf{E} \tr \{ U^\dagger_j \,
r^{(\tau)}_j{}^\dagger \delta\!H r^{(\tau)}_j \, U_j \,
r^{(\tau)}_j{}^\dagger \delta\!H r^{(\tau)}_j \} \\
+ \mathcal{O}(\delta^3).
\end{multline}

%
\bibliographystyle{apsrev}
\bibliography{references}
%
%
%

\end{document}